# Time complexity of in-memory solution of linear systems

Zhong Sun, *Member, IEEE*, Giacomo Pedretti, Piergiulio Mannocci, Elia Ambrosi,
Alessandro Bricalli, and Daniele Ielmini, *Fellow, IEEE*

*Abstract*—In-memory computing with crosspoint resistive memory arrays has been shown to accelerate data-centric computations such as the training and inference of deep neural networks, thanks to the high parallelism endowed by physical rules in the electrical circuits. By connecting crosspoint arrays with negative feedback amplifiers, it is possible to solve linear algebraic problems such as linear systems and matrix eigenvectors in just one step. Based on the theory of feedback circuits, we study the dynamics of the solution of linear systems within a memory array, showing that the time complexity of the solution is free of any direct dependence on the problem size $N$, rather it is governed by the minimal eigenvalue of an associated matrix of the coefficient matrix. We show that, when the linear system is modeled by a covariance matrix, the time complexity is $O(\log N)$ or $O(1)$. In the case of sparse positive-definite linear systems, the time complexity is solely determined by the minimal eigenvalue of the coefficient matrix. These results demonstrate the high speed of the circuit for solving linear systems in a wide range of applications, thus supporting in-memory computing as a strong candidate for future big data and machine learning accelerators.

## I. INTRODUCTION

THE system of linear equations is among the most common problems in scientific and engineering fields, such as quantum mechanics, statistical analysis, network theory and machine learning [1], [2]. Improving time and energy efficiencies of solving linear systems is constantly sought in modern scientific computing [3] and data-centric applications [4]. Conventional digital computers solve linear systems by using classical algorithms such as Gaussian elimination, LU factorization and conjugate gradient (CG) method [5]. In these algorithms, the time complexity is always a polynomial function of matrix size $N$, namely $O(\text{poly}(N))$. In the era of big data and internet of things, however, such performance may not be sufficient, given the exponential increase of data size and the approaching physical limits of the Moore's law [6]. In the quest for an acceleration of data-intensive tasks, quantum computing has also been demonstrated to solve systems of linear equations with an $O(\log N)$ time complexity [7], [8]. Although quantum computing appears promising for exponential speedup of the solution, cryogenic temperatures and maintenance of quantum coherence in quantum computers appear as strong obstacles toward practical implementation especially for portable computing [9]. Here we show that in-memory computing, which relies on the physical computing with crosspoint analog resistive memory arrays and negative feedback in circuit connections, solves a linear system in a time that is dictated by the minimal eigenvalue of an associated matrix. As a result, the corresponding time complexity is demonstrated to be extremely low, *e.g.*, $O(\log N)$ or $O(1)$ for solving linear systems of model covariance matrix. For sparse positive-definite linear systems, the time complexity depends solely on the minimal eigenvalue of the coefficient matrix, thus outperforming the conventional digital and quantum computing counterparts.

## II. MULTILEVEL RRAM DEVICE

Resistive memories (also known as memristors) are two-terminal devices whose resistance (conductance) can be changed by a voltage stimulus [10], [11]. The class of resistive memory devices includes various concepts, such as the resistive switching memory (RRAM) [12], [13], the phase change memory (PCM) [14] and the magneto-resistive memory (MRAM) [15]. Thanks to their small size and nonvolatile behavior, resistive memories have been widely considered as promising devices for memory technology [12], [13]. Most importantly, resistive memories enable stateful logic [16], [17] and in-memory analog computing [18], [19], thus circumventing the communication bottleneck between the memory and the processor which represents the main limitation of von Neumann machines. Fig. 1 shows the multilevel current-voltage (*I-V*) characteristics of a RRAM device, supporting the ability to store an arbitrary analog number mapped in the device conductance [19]-[21]. The RRAM conductance is controlled by the compliance current, namely the maximum current supplied by the select transistor during the set transition from high resistance to low resistance [17]. The device is fully reconfigurable, in that the application of a negative voltage can restore a high resistance in the device, thus preparing for another analog set operation. The 8

This work has received funding from the European Research Council (ERC) under the European Union's Horizon 2020 Research and Innovation Programme (grant agreement No. 648635). This work was partially carried out at Polifab, the micro- and nanofabrication facility of Politecnico di Milano. (*Corresponding author: Zhong Sun, and Daniele Ielmini.*)

Z. Sun, G. Pedretti, P. Mannocci, E. Ambrosi, A. Bricalli, and D. Ielmini are with the Dipartimento di Elettronica, Informazione e Bioingegneria, Politecnico di Milano, 20133 Milan, Italy (e-mail: zhong.sun@polimi.it, daniele.ielmini@polimi.it).

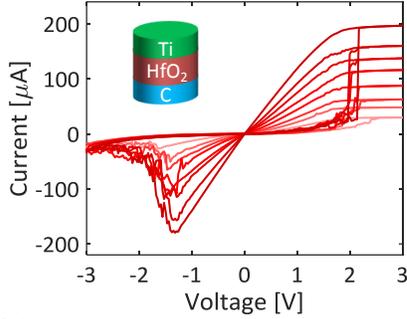

Fig. 1. *I-V* characteristics of multilevel operations of the RRAM device. 8 conductance levels are shown, and the values read at a small voltage are 120, 80, 60, 50, 30, 20, 15 and 10 $\mu S$, respectively. The inset shows the RRAM device structure.

conductance levels in Fig. 1 will be employed in the following as discrete values to construct matrices and simulate the solution of linear systems within the circuit.

## III. TIME COMPLEXITY OF SOLVING LINEAR SYSTEMS

### A. Time Complexity Analysis of the Crosspoint Circuit

Crosspoint resistive memory arrays can be conveniently used to accelerate the matrix-vector multiplication (MVM), which is a core operation in many computing tasks, such as sparse coding [18], signal processing [19] and neural network training [22]. Recently, a crosspoint circuit of resistive memory arrays have been demonstrated to solve linear systems or eigenvector equations in one step [23]. Fig. 2(a) shows the circuit to solve a system of linear equations, which reads

$$\mathbf{A}\mathbf{x} = \mathbf{b}, \tag{1}$$

where $\mathbf{A}$ is an $N \times N$ matrix of coefficients, $\mathbf{b}$ is a known vector, and $\mathbf{x}$ is the unknown vector to be solved. In the crosspoint circuit, each coefficient $A_{ij}$ of matrix $\mathbf{A}$ is coded as the analog conductance $G_{ij}$ of a resistive memory, the input voltages represents $-\mathbf{b}$, and the output voltages of the operational amplifiers (OAs) provides the solution $\mathbf{x} = \mathbf{A}^{-1}\mathbf{b}$. The reconfigurable resistive memory enables the crosspoint circuit of Fig. 2(a) to map an arbitrary matrix $\mathbf{A}$ with positive coefficients.

To address the time complexity of the crosspoint circuit, namely the time it takes to yield the correct answer to the problem, we first note that the closed feedback loop plays a leading role in ensuring a physical iteration between input and output. In other words, instead of completing a certain number of open loop iterations with a gradually diminishing error, we let the signal physically circulate within the closed loop to minimize the error in the feedback network, thus enabling a virtually instantaneous solution. In reality, the non-idealities of the circuit, such as the limited response time of the OAs, result in a finite time complexity of the solution.

Fig. 2(b) shows a block diagram of the crosspoint circuit, where the crosspoint array plays the role of feedback network conveying the weighted output to be compared with the input, thus establishing a stable output. To study the time response of the circuit, we write the input-output relationship in terms of

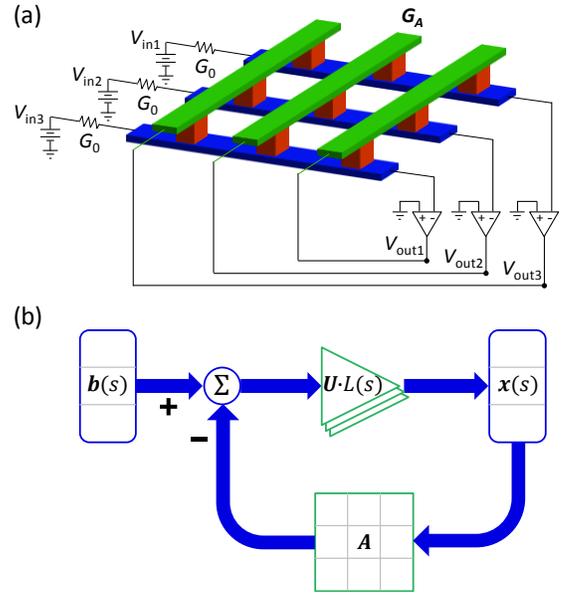

Fig. 2. (a) Crosspoint resistive memory circuit for solving linear systems, illustrated with $N = 3$ as the problem size. The conductance matrix $\mathbf{G_A}$ maps $\mathbf{A}$, the input voltages $[V_{in1}; V_{in2}; V_{in3}]$ represents $-\mathbf{b}$, and the output voltages $[V_{out1}; V_{out2}; V_{out3}]$ gives the solution of $\mathbf{x}$. (b) Block diagram of the crosspoint circuit as a control system. The crosspoint array conveys the output $\mathbf{x}$ to interact with the input $\mathbf{b}$.

Laplace transform of the *i*th OA in Fig. 2(a), according to the Kirchhoff's voltage and OA theory, namely:

$$\frac{\sum_j G_{ij} V_{out,j}(s) + G_0 V_{in,i}(s)}{\sum_i G_{ij} + G_0} L(s) = V_{out,i}(s), \tag{2}$$

where $G_{ij}$ is the conductance of the *j*th device in the *i*th row, $G_0$ is the input conductance, $L(s)$ is the open-loop gain of the OA, $s$ is the complex variable in Laplace transform. The ratio between $G_{ij}$ and $G_0$ gives the corresponding element of matrix $A$, namely $A_{ij} = G_{ij}/G_0$. Replacing $V_{out}$ and $V_{in}$ with $x$ and $-b$ respectively results in the following equation:

$$-\frac{1}{1+\sum_j A_{ij}}\left[\sum_j A_{ij} x_j(s) - b_i(s)\right]L(s) = x_i(s). \tag{3}$$

For the whole system, all equations can be combined in the form of a matrix equation, namely

$$-\mathbf{U}[\mathbf{A}\mathbf{x}(s) - \mathbf{b}(s)]L(s) = \mathbf{x}(s), \tag{4}$$

where $\mathbf{U}$ is a diagonal matrix defined as $\mathbf{U} = diag\left(\frac{1}{1+\sum_j A_{1j}}, \frac{1}{1+\sum_j A_{2j}}, \cdots, \frac{1}{1+\sum_j A_{Nj}}\right)$. As all OAs in the circuit are assumed identical, $L(s)$ is a scalar linking the inverting input and the output of each OA. Assuming a single-pole transfer function [24] for the employed OAs, namely $L(s) = L_0/\left(1 + \frac{s}{\omega_0}\right)$, where $L_0$ is the DC open-loop gain and $\omega_0$ is the 3-dB bandwidth, Eq. (4) becomes

$$s\mathbf{x}(s) = -L_0\omega_0\left[\left(\mathbf{M} + \frac{1}{L_0}\mathbf{I}\right)\mathbf{x}(s) - \mathbf{U}\mathbf{b}(s)\right], \tag{5}$$

where $\mathbf{M} = \mathbf{U}\mathbf{A}$ is matrix associated with matrix $\mathbf{A}$, $\mathbf{I}$ is the $N \times N$ identity matrix. As $L_0$ is usually much larger than 1, the second term in $\mathbf{M} + \frac{1}{L_0}\mathbf{I}$ can be omitted. As a result, the inverse Laplace transform of Eq. (5) into the time domain gives the differential equation:

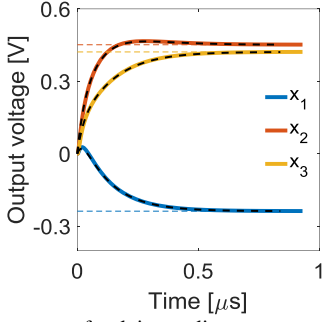

Fig. 3. Time response of solving a linear system with the circuit. The conductance matrix is $G_A = [120,15,80; 50,50,60; 60,10,80]$ $\mu$S, and $G_0 = 100$ $\mu$S. The input vector is $b = -[0.12; 0.36; 0.24]$. Colored full lines: transient curves in SPICE simulation. Colored dash lines: analytical solutions. Black dash lines: simulated time response with the FD algorithm.

$$\frac{dx(t)}{dt} = -L_0\omega_0[Mx(t) - Ub(t)], \quad (6)$$

which describes the dynamics of the crosspoint circuit for solving Eq. (1). Though an analytical solution can be obtained for Eq. (6), we developed an iterative algorithm to analyze the transient behavior and evaluate the time complexity of the crosspoint circuit. Eq. (6) can be approximated by a finite difference (FD) equation, namely:

$$x(t + \Delta t) = \alpha Ub + (I - \alpha M)x(t), \quad (7)$$

where $\alpha$ is a small dimensionless number given by $\alpha = L_0\omega_0\Delta t$ with $\Delta t$ being the incremental time. To verify the FD algorithm, we have run transient simulation for solving a linear system, comparing the iterative solution according to Eq. (7) with the SPICE (simulation program with integrated circuit emphasis) transient simulation result. A 3×3 matrix was randomly constructed with the 8 discrete values in Fig. 1, and the corresponding linear system was solved. Fig. 3 shows the time evolution of the output $x(t)$ for the linear system. The trajectories of FD algorithm results appear highly consistent with the ones of circuit simulation, also both the asymptotic results are in line with the steady-state solution. A concern about the OAs is the slew rate, which limits the response time of the output in case of large signals. Our adopted OA (AD823 from Analog Devices) has a slew rate of 22 V/$\mu$s, which guarantees that the circuit operates in the small signal response area in our simulation. From the simulation results, the steady-state output amplitude is reached in a computing time below 1 $\mu$s, which is defined as the time for the norm of error dropping below $10^{-3}$.

The convergence of the iterative algorithm in Eq. (7) requires that the spectral radius of the matrix $I - \alpha M$ has to be less than 1, which implies that the minimal eigenvalue (or real part of eigenvalue) $\lambda_{M,min}$ of the associated matrix $M$ has to be positive. The $\lambda_{M,min}$ condition can also be understood from the viewpoint of transfer function of the circuit, which is $T(s) = -\left(M + \frac{s}{L_0\omega_0}I\right)^{-1} U$ according to Eq. (5). The poles of the system can be determined by assigning $M + \frac{s}{L_0\omega_0}I$ as a singular matrix, which implies that the poles are located at $s = -L_0\omega_0\lambda_M$, where $\lambda_M$ is an eigenvalue of matrix $M$. For the system to be stable, the $N$ $\lambda_M$'s (or their real parts) have all to be positive [25]. As $U$ is a positive diagonal matrix, the $\lambda_{M,min}$ condition is conveniently satisfied by positive-definite (PD) matrix, which is widely encountered in various fields and applications, such as statistical analysis [26], quantum chemistry simulation [27] and network theory [28]. For this reason, we shall focus our attention on PD matrix in the following.

To provide an analytical model for the computing time as a function of $\lambda_{M,min}$, we have analyzed the convergence behavior of the iterative algorithm. The convergence is measured in the A-norm, which is defined as $\|x\|_A = \sqrt{x^T A x}$ [5]. In the case of positive definite matrix $A$, there is $\|x\|_A = \left\|A^{\frac{1}{2}}x\right\|_2$, where $\|\cdot\|_2$ is the $\ell_2$ norm. Similarly, the induced matrix norm follows $\|B\|_A = \left\|A^{\frac{1}{2}}BA^{-\frac{1}{2}}\right\|_2$. If a linear system is solved with a precision $\epsilon$ at time $t$, the A-norm of solution error has to satisfy

$$\|x(t) - x^*\|_A \leq \epsilon, \quad (8)$$

where $x^* = A^{-1}b$ is the precise solution. $\|x(t) - x^*\|_A$ follows

$$\begin{aligned}\|x(t) - x^*\|_A^2 &= \|\alpha Ub + (I - \alpha M)x(t - \Delta t) - x^*\|_A^2 \\ &= \|(I - \alpha M)x(t - \Delta t) - (I - \alpha M)x^*\|_A^2 \\ &\leq \|I - \alpha M\|_A^2 \|x(t - \Delta t) - x^*\|_A^2 \\ &= \left\|A^{\frac{1}{2}}(I - \alpha UA)A^{-\frac{1}{2}}\right\|_2^2 \|x(t - \Delta t) - x^*\|_A^2 \\ &= \left\|I - \alpha A^{\frac{1}{2}}UA^{\frac{1}{2}}\right\|_2^2 \|x(t - \Delta t) - x^*\|_A^2 \quad (9)\end{aligned}$$

By defining $W = A^{\frac{1}{2}}UA^{\frac{1}{2}}$, which is a positive definite matrix, there is $\|I - \alpha W\|_2 = 1 - \alpha\lambda_{W,min}$ with $\lambda_{W,min}$ being the minimal eigenvalue of $W$. Matrix $M$ has the same eigenvalues as $W$, so the inequality of Eq. (9) becomes

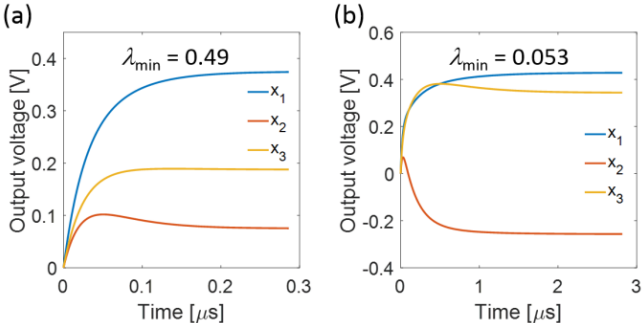

Fig. 4. (a) Transient behavior of solving a linear system of a 3×3 PD matrix with a relatively large $\lambda_{min}$, which is labeled on the top. (b) Same as (a), but for a matrix with a one-order smaller $\lambda_{min}$.

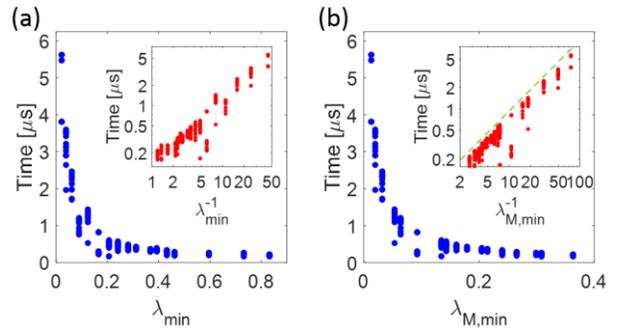

Fig. 5. (a) Summary of computing time for solving linear systems with different $\lambda_{min}$'s. The inset shows the computing time as a function of $1/\lambda_{min}$. (b) Computing time as a function of $\lambda_{M,min}$. The inset shows the computing time as a function of $1/\lambda_{M,min}$, indicating a precise linear upper bound (green line).

$$\|x(t) - x^*\|_A^2 \leq (1 - \alpha\lambda_{M,min})^2 \|x(t - \Delta t) - x^*\|_A^2 \quad (10)$$

Repeating the above process iteratively leads to

$$\|x(t) - x^*\|_A^2 \leq (1 - \alpha\lambda_{M,min})^{2t/\Delta t} \|x(0) - x^*\|_A^2$$
$$< e^{-2\lambda_{M,min}L_0\omega_0 t} \|x^*\|_A^2$$
$$= e^{-2\lambda_{M,min}L_0\omega_0 t} x^{*T}b \quad (11)$$

The upper bound of $\|x(t) - x^*\|_A$ satisfying inequation (8) finally reveals the computing time as

$$\tau = \frac{1}{\lambda_{M,min}L_0\omega_0} \log\frac{\sqrt{x^{*T}b}}{\epsilon}. \quad (12)$$

Note that the inner product $x^{*T}b$ of input and output is always positive by the definition of PD matrix. Also, compared with the reciprocal impact of $\lambda_{M,min}$ on the computing time, the logarithmic role of $x^{*T}b$ is suppressed. Therefore, the time complexity for solving linear systems with the crosspoint circuit is $O\left(\frac{1}{\lambda_{M,min}} \log\frac{1}{\epsilon}\right)$, which shows no direct dependence on the matrix size $N$. The time complexity of conventional iterative algorithms is usually a polynomial function of $N$, with also the matrix properties such as eigenvalues involved [2]. We write the time complexity of the crosspoint circuit in a similar form, by linking $\lambda_{M,min}$ to the minimal eigenvalue $\lambda_{min}$ of matrix $A$, namely $\lambda_{M,min} = u\lambda_{min}$. Therefore, the time complexity is $O\left(\frac{1}{u\lambda_{min}} \log\frac{1}{\epsilon}\right)$, where the critical role of $\lambda_{min}$ is recognized, and the factor $u$ may contribute an $N$-dependence.

To support the $\lambda_{min}$-controlled time complexity of the crosspoint circuit, we considered two 3×3 PD matrices containing discrete conductance levels in Fig. 1 with $\lambda_{min} = 0.49$ and $0.053$, respectively. Fig. 4 shows the SPICE transient responses for the two linear systems. The simulation results indicate a faster response for the larger $\lambda_{min}$, thus supporting the dominant role of $\lambda_{min}$. Fig. 5(a) summarizes the computing times for various 3×3 PD matrices, spanning two decades of $\lambda_{min}$, and assuming 15 random vectors $b$ for each matrix $A$. The results show that the computing time is inversely proportional to $\lambda_{min}$, as also supported by the plot of computing time as a function of $1/\lambda_{min}$ in the inset. As can be observed, there is a rough upper bound for the computing time, which scales linearly with $1/\lambda_{min}$ and defines the time complexity of solving linear systems. In Fig.

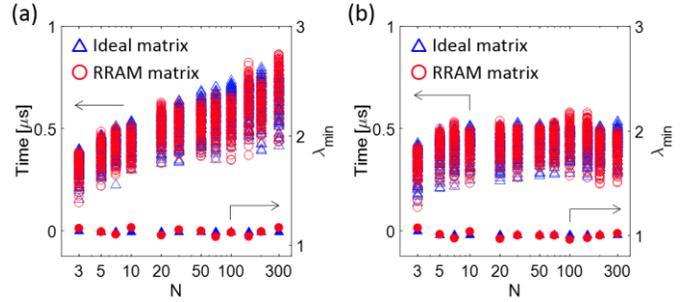

Fig. 7. (a) Summary of computing time for solving linear systems of the first-order model covariance matrix with different sizes, ranging from 3×3 to 300×300. Results from both ideal matrix and RRAM conductance matrix are shown. The $\lambda_{min}$'s of both ideal matrix and RRAM matrix are shown as the right $y$ axis. (b) Same as (a), but for the second-order model covariance matrix.

5(b), the computing time show a precise linearity for the upper bound with the increase of $1/\lambda_{M,min}$, demonstrating the precise description of time complexity by Eq. (12).

### B. Time Complexity of Solving Model Linear Systems

To show the time complexity dependence on the matrix size $N$, we considered a model covariance matrix to represent a real-world problem [4], [29], [30], namely:

$$A_{ij} = \begin{cases} \frac{1}{|i-j|^\beta}, & \text{if } i \neq j \\ 1 + \sqrt{i}, & \text{if } i = j \end{cases}, \quad (13)$$

where $\beta > 0$ is a decay factor. Covariance matrix plays an important role in statistical inference and financial economics, such as in the portfolio theory. The decrease of off-the-diagonal elements of the matrix was chosen to simulate the decreasing correlation of high-dimensional data samples in a realistic covariance matrix. In the following, we consider model covariance matrices of the first order ($\beta = 1$) and the second order ($\beta = 2$). Note that $\lambda_{min}$ is asymptotically constant as the size of the model covariance matrix increases, thus the $N$-dependence of time complexity is related solely to the factor $u$.

In simulating the solution of a linear system of a model covariance matrix, the coefficients in Eq. (13) were mapped in the crosspoint array with 64 discrete and uniform conductance levels, which is feasible for previously reported resistive memories [19], [31]-[33]. The conductance ratio, defined as the ratio between the maximum conductance $G_{max}$ and minimum conductance $G_{min}$, was assumed equal to $10^3$, which is also achievable for various RRAM devices [34], [35]. Each level was randomized according to normal distribution with a standard deviation $\sigma = \Delta G/6$, where $\Delta G = G_{max}/64$ is the nominal difference between two adjacent conductance levels.

Fig. 6(a) shows the crosspoint conductance for a 10×10 first-order covariance matrix implemented with RRAM devices. We simulated matrix inversion by the crosspoint circuit, which is equivalent to solving $N$ linear systems where the input vector $b$ is set equal to each column of the identity matrix [23]. Fig. 6(b) shows the 100 computed elements of $A^{-1}$ as a function of the analytical results, which indicates a good accuracy.

To study the scaling behavior of the computing time, linear systems were solved for matrix size ranging from $N = 3$ to 300. For each matrix $\boldsymbol{A}$, 100 linear systems were solved with random input vectors $\boldsymbol{b}$. Fig. 7(a) shows the simulated computing time for the first-order covariance matrix. The results reveal that the computing time scales logarithmically with the matrix size $N$, i.e., the time complexity is $O(\log N)$. The $O(\log N)$ time complexity indicates the coefficient $u$ scales as $u \propto \frac{1}{\ln N}$. The figure also shows the analytical minimal eigenvalues and those calculated for the conductance matrices implemented in the crosspoint circuit. The difference between the analytical and calculated eigenvalues due to conductance discretization and randomization is responsible for the inconsistency of the computing times obtained by the ideal and conductance matrices. This interpretation is supported by the fact that, for instance, the computing time for the crosspoint resistive-memory simulation with $N = 10$ is smaller than the ideal value, while the minimal eigenvalue is larger. The opposite case applies to $N = 150$. To guarantee the minimal eigenvalue is in the vicinity of the ideal one and thus the computing time is predictable, it is important to reduce device variations by using devices of large conductance window accommodating sufficient analog levels, also by using verify algorithms for device programming [20].

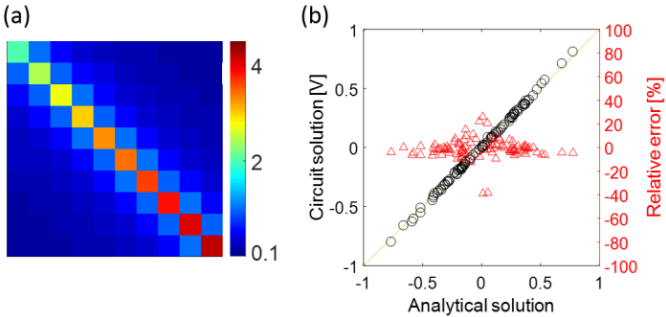

Fig. 6. (a) The 10×10 first-order model covariance matrix mapped by discretized and randomized RRAM devices. The conductance unit is 100 $\mu$S. (b) The inverse matrix solved with the crosspoint RRAM circuit, as a function of the precise analytical solution. The relative errors (right $y$ axis) are generally small, except for the entries near close to zero.

Fig. 7(b) shows the scaling behavior of computing time for the second-order covariance matrix, indicating a constant computing time, i.e., the time complexity is $O(1)$. Due to the strong decaying behavior, the elements far from the diagonal are close to zero, thus requiring a larger conductance ratio ($G_{max}/G_{min} = 10^4$) of resistive memories to map the entire matrix. The $O(1)$ time complexity in Fig. 7(b) reveals the coefficient $u$ is asymptotically constant for the second-order covariance matrix. Therefore, depending on the matrix structure, extremely low time complexity such as $O(\log N)$ or $O(1)$ can be achieved, which hugely reduces the computing time for large-scale problems.

*C. Comparison with Other Computing Paradigms*

The quantum algorithm for solving linear systems addresses sparse Hermitian matrix, and its time complexity is $O\left(\frac{s^2 \lambda_{max}^2}{\epsilon \lambda_{min}^2} \log N\right)$, where $\lambda_{max}$ is the maximum eigenvalue of the matrix, $\lambda_{max}/\lambda_{min}$ is the condition number, and $s$ is the sparsity which means the matrix has at most $s$ nonzero entries per row [7]. To make a direct comparison with quantum computing, we also consider sparse PD matrix that is a subset of real-valued Hermitian matrix. By defining $U_{min}$ as the minimal eigenvalue (also the minimal element) of the diagonal matrix $\boldsymbol{U}$, there is a relation $\lambda_{M,min} \geq U_{min}\lambda_{min}$, due to the eigenvalue inequality for a matrix product [36]. As $U_{min}$ is determined by the inverse of the largest row sum of matrix $A$, there is $\frac{1}{\lambda_{M,min}} \leq \frac{1}{U_{min}\lambda_{min}} \sim \frac{s}{\lambda_{min}}$. As a result, the time complexity of the crosspoint circuit in Eq. (12) is reduced as $O\left(\frac{s}{\lambda_{min}} \log \frac{1}{\epsilon}\right)$, or $O\left(\frac{1}{\lambda_{min}}\right)$ in line with the $O\left(\frac{\lambda_{max}^2}{\lambda_{min}^2} \log N\right)$ time complexity of the quantum algorithm.

We tested a set of sparse PD linear systems to verify the time complexity of the crosspoint circuit, with the sparsity assumed as $s = 10$. We generated 1,000 linear systems, i.e. 1,000 sparse PD matrices and one random input vector for each, with sizes from 20×20 to 200×200. Fig. 8(a) shows the computing time for solving the 1,000 linear systems, which is independent of $N$ and is solely determined by $\lambda_{min}$, thus supporting the $O\left(\frac{1}{\lambda_{min}}\right)$ time complexity. Fig. 8(b) shows the computing time for a subset of the 1,000 linear systems with limited $\lambda_{min}$ to exclude its contribution. The relative computing time of the quantum algorithm for solving the same linear systems is also shown, according to its time complexity formula. Fig. 8(b) also reports the relative computing time of the CG method [37], which is the most efficient algorithm for solving PD linear systems in conventional digital computers thanks to a time complexity of $O\left(Ns\sqrt{\frac{\lambda_{max}}{\lambda_{min}}} \log \frac{1}{\epsilon}\right)$ for sparse matrix. The results indicate that in-memory computing, quantum computing and digital computing display $O(1)$, $O(\log N)$ and $O(N)$ time complexities, respectively, for solving sparse PD linear systems.

## IV. DISCUSSION

The analysis of circuit dynamics and time complexity is based on the assumption of an ideal crosspoint resistive array, as the RC delay in crosspoint MVM is extremely low, e.g., 0.5 nanosecond in a 1024×1024 array [38]. To evaluate the impact of wire resistance, parasitic capacitance and device capacitance on time complexity of the circuit, we have simulated the solution of linear systems of the model covariance matrix in SPICE with these parasitic components considered. Specifically, each crosspoint resistive device was replaced by a sub-circuit module (Fig. 9), where the wire resistance and the parasitic capacitance are assumed according to interconnect parameters at 65 nm node in the ITRS table [39], and the device capacitance is calculated with dielectric constant of $HfO_2$ [40]. The simulation results of the solution of linear systems for increasing size is reported in Fig. 10. The results indicate the same time complexity as the ideal circuit, namely $O(\log N)$ and $O(1)$ for solving linear systems of the first-order and the second-order model variance matrix, respectively. Such a comparison supports the robustness of the crosspoint computing circuit against parasitics. The wire resistance imposes a relatively small error to the steady-state solution, which is alleviated by the intermediate interconnect technology for crosspoint arrays, in contrast to the aggressive downscaling of conventional high-density memory [38]. Also, increasing the crosspoint device resistance and adopting 3D integration are helpful to improving the solution accuracy [23].

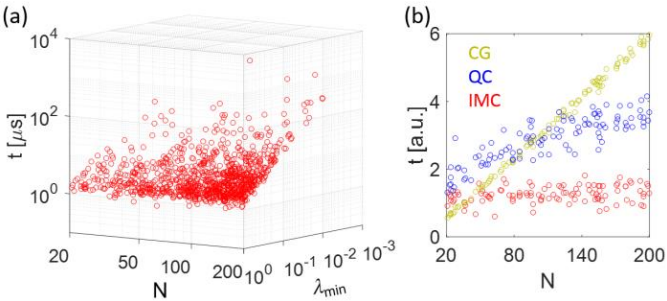

Fig. 8. (a) Summary of computing time for solving 1,000 sparse PD linear systems, plotted as a function of $N$ and $\lambda_{min}$. (b) A subset of simulation results for $\lambda_{min}$ limited within [0.9, 1], showing a constant computing time for in-memory computing (IMC). The relative computing time of conventional CG method and quantum computing (QC) for solving these linear systems are also calculated, according to their time complexity formulas.

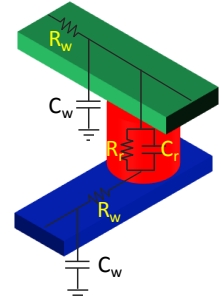

Fig. 9. Sub-circuit module of a single crosspoint resistive memory device. $R_r$ is device resistance, storing an element value in the matrix, $C_r$ is device capacitance, $R_w$ is wire resistance, $C_w$ is parasitic capacitance.

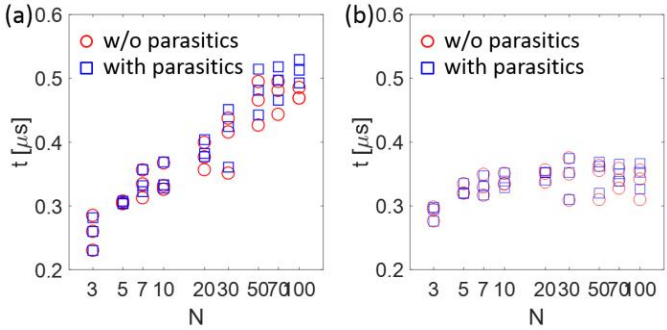

Fig. 10. Time complexity of the crosspoint circuit without or with parasitics for (a) the first-order and (b) the second-order model covariance matrix. The solution precision $\epsilon$ is limited to $10^{-2}$ due to the more discrete outputs when approaching the steady state in SPICE, thus the computing time is less than the ones in Fig. 7. Also, the matrix was limited to $100 \times 100$ for the circuit complexity consideration with parasitic components.

In conventional computers, linear systems of a dense matrix can be solved with standard algorithms such as Gaussian elimination and LU factorization, which are of $O(\text{poly}(N))$ time complexities. The solution can be accelerated with parallel algorithms, for instance, Gaussian elimination can be carried out in parallel with a time complexity of $O(N)$ by using $N^2$ processors [41]. Csanky's algorithm reports a better time complexity that is $O(\log^2 N)$, while $N^4$ processors are required [42]. In the crosspoint computing circuit, $O\left(\frac{1}{\lambda_{M,min}}\right)$ time complexity that may implicate $O(\log N)$ or even $O(1)$ is achieved with only $N^2$ memory devices to necessarily implement the matrix, thus representing a much more efficient method for solving linear systems. Note that the $N \times N$ crosspoint array is imperative to store the matrix, and the data are processed directly in the memory, whereas in conventional computers, the memory cost also scales as $N^2$, and an amount of additional processors are required. Compared with in-memory computing, digital computing possesses an additional data access complexity due to the communication between the separated memory and processor [43]. Therefore, there is an obvious efficiency advantage for solving linear systems with the crosspoint resistive memory circuit, thanks to the concept of in-memory computing and to the unique time complexity of computation. In the linear system problem, different matrices may be involved to be stored in the crosspoint array, thus requiring device reprogramming. In this sense, fast and reliable writing schemes [21] are favored to retain the advantage of in-memory computing. The high efficiency of our method is attributed to the parallelism in the circuit, where the Kirchhoff's voltage law and the concurrent feedback play major roles. According to the output update algorithm in Eq. (7), the whole system resembles the Hopfield network [44], [45], which is well known for its physics-inspired high parallelism. In contrast, there is no discrete iteration in the crosspoint circuit, instead the output evolves in a self-sustained fashion, thus contributing to an even higher speed in addition to the architecture parallelism.

## V. Conclusion

In conclusion, we have studied the time complexity of solving linear systems with an in-memory computing circuit. Based on the feedback theory, we show that only if the minimal eigenvalue (or real part of eigenvalue) $\lambda_{M,min}$ of the associated matrix is positive, the linear system can be solved by the circuit. According to a finite difference algorithm developed for the circuit dynamic, we show that the time complexity is free of direct $N$-dependence, rather determined solely by $\lambda_{M,min}$. For solving linear systems where $\lambda_{M,min}$ possesses a weak (or none) $N$-dependence, the speed of the circuit is expected to be unprecedentedly high, *e.g.*, the time complexity is $O(\log N)$ or $O(1)$ for solving linear systems of the model covariance matrices. When addressing sparse PD linear systems, the time complexity is $O\left(\frac{1}{\lambda_{min}}\right)$, thus outperforming its counterparts of conventional digital computing and quantum computing. We project that, when analog non-volatile memory technology becomes maturely industrialized, in-memory computing can play a leading role in boosting the computing performance for big data in a wide range of real-world applications.

## References

[1] C. D. Meyer, *Matrix Analysis and Applied Linear Algebra*. Philadelphia, PA, USA: SIAM, 2000.
[2] Y. Saad, *Iterative Methods for Sparse Linear Systems*, 2nd ed. Philadelphia, PA, USA: SIAM, 2003.
[3] K. Bourzac, "Stretching supercomputers to the limit", *Nature*, vol. 551, pp. 554-556, () 2017.
[4] M. Le Gallo *et al.*, "Mixed-precision in-memory computing", *Nat. Electron.*, vol. 1, no. 4, pp. 246-253, Apr. 2018.
[5] G. H. Golub, C. F. van Loan, *Matrix Computations*, 4th ed. Baltimore, MD, USA: Johns Hopkins University Press, 2013.
[6] M. M. Waldrop, "The chips are down for Moore's law," *Nature*, vol. 530, pp. 144–147, Feb. 2016
[7] A. W. Harrow, A. Hassidim, S. Lloyd, "Quantum Algorithm for Linear Systems of Equations," *Phys. Rev. Lett.*, vol. 103, pp. 150502, Oct. 2009.
[8] Y. Zheng *et al.*, "Solving Systems of Linear Equations with a Superconducting Quantum Processor," *Phys. Rev. Lett.*, vol. 118, pp. 210504, May 2017.

[9] T. D. Ladd, F. Jelezko, R. Laflamme, Y. Nakamura, C. Monroe, J. L. O'Brien, "Quantum computers," *Nature*, vol. 464, pp. 45-53, Mar. 2010.

[10] R. Waser, R. Dittmann, G. Staikov, K. Szot, "Redox-Based Resistive Switching Memories - Nanoionic Mechanisms, Prospects, and Challenges," *Adv. Mater.*, vol. 21, pp. 2632-2663, Jul. 2009.

[11] D. Ielmini, R. Waser, *Resistive Switching: From Fundamentals of Nanoionic Redox Processes to Memristive Device Applications*. Hoboken, NJ, USA: Wiley-VCH, 2015.

[12] H.-S. P. Wong *et al.*, "Metal-Oxide RRAM," *Proc. IEEE*, vol. 100, pp. 1951-1970, Jun. 2012.

[13] D. Ielmini, "Resistive switching memories based on metal oxides: Mechanisms reliability and scaling," *Semicond. Sci. Technol.*, vol. 31, pp. 063002, May 2016.

[14] S. Raoux, W. Welnic, D. Ielmini, "Phase change materials and their application to non-volatile memories," *Chem. Rev.*, vol. 110, pp. 240-267, 2010.

[15] A. D. Kent, D. Worledge, "A new spin on magnetic memories," *Nat. Nanotechnol.*, vol. 10, pp. 187-191, Mar. 2015.

[16] J. Borghetti, G. S. Snider, P. J. Kuekes, J. J. Yang, D. R. Stewart, R. S. Williams, "'Memristive' switches enable 'stateful' logic operations via material implication," *Nature*, vol. 464, pp. 873-876, Apr. 2010.

[17] Z. Sun, E. Ambrosi, A. Bricalli, D. Ielmini, "Logic computing with stateful neural networks of resistive switches," *Adv. Mater.*, vol. 30, no. 38, pp. 1802554, Sep. 2018.

[18] P. M. Sheridan, F. Cai, C. Du, W. Ma, Z. Zhang, W. D. Lu, "Sparse coding with memristor networks," *Nat. Nanotechnol.*, vol. 12, pp. 784-789, May 2017.

[19] C. Li *et al.*, "Analogue signal and image processing with large memristor crossbars," *Nat. Electron.*, vol. 1, pp. 52-59, Dec. 2017.

[20] F. Alibart, L. Gao, B. D. Hoskins, D. B. Strukov, "High precision tuning of state for memristive devices by adaptable variation-tolerant algorithm," *Nanotechnology*, vol. 23, pp. 075201, Jan. 2012.

[21] C. Li *et al.*, "Efficient and self-adaptive in-situ learning in multilayer memristor neural networks," *Nat. Commun.*, vol. 9, pp. 2385, Jun. 2018.

[22] G. W. Burr *et al.*, "Experimental demonstration and tolerancing of a large-scale neural network (165 000 synapses) using phase-change memory as the synaptic weight element," *IEEE Trans. Electron Devices*, vol. 62, pp. 3498-3507, Nov. 2015.

[23] Z. Sun, G. Pedretti, E. Ambrosi, A. Bricalli, D. Ielmini, "Solving matrix equations in one step with cross-point resistive arrays," *Proc. Natl. Acad. Sci. USA*, vol. 116, no. 10, pp. 4123-4128, Mar. 2019.

[24] B. Razavi, *Design of Analog CMOS Integrated Circuits*. New York, NY, USA: McGraw-Hill, 2001.

[25] C. Moler, C. Van Loan, "Nineteen Dubious Ways to Compute the Exponential of a Matrix, Twenty-Five Years Later," *SIAM Rev.*, vol. 45, pp. 3-49, 2003.

[26] R. Bhatia, *Positive definite matrices*. Princeton, NJ, USA: Princeton University Press, 2007.

[27] R. G. Parr, W. Yang, *Density-Functional Theory of Atoms and Molecules, vol. 16*. Oxford, UK: Oxford University Press, 1989.

[28] R. Rojas, *Neural Networks: A Systematic Introduction*. Berlin, Germany: Springer-Verlag, 1996.

[29] C. Bekas, A. Curioni, I. Fedulova, (2009) "Low cost high performance uncertainty quantification," in *Proc. 2nd Workshop on High Performance Computational Finance*, Portland, OR, USA, 2009, pp. 8:1-8:8.

[30] J. Tang, Y. Saad, "A probing method for computing the diagonal of a matrix inverse," *Numer. Linear Algebra Appl.*, vol. 19, pp. 485-501, May 2012.

[31] K. Seo *et al.*, "Analog memory and spike-timing-dependent plasticity characteristics of a nanoscale titanium oxide bilayer resistive switching device", *Nanotechnology*, vol. 22, pp. 254023, May 2011.

[32] J. Park, M. Kwak, K. Moon, J. Woo, D. Lee, H. Hwang, "$TiO_x$-based RRAM Synapse with 64-levels of Conductance and Symmetric Conductance Change by Adopting a Hybrid Pulse Scheme for Neuromorphic Computing," *IEEE Electron Device Lett.*, vol. 37, pp. 1559-1562, Dec. 2016.

[33] J. Tang *et al.*, "ECRAM as Scalable Synaptic Cell for High-Speed, Low-Power Neuromorphic Computing," *IEEE International Electron Devices Meeting*, pp. 13.1.1-13.1.4, Dec. 2018.

[34] T.-C. Chang, K.-C. Chang, T.-M. Tsai, T.-J. Chu, S. M. Sze, "Resistance random access memory," *Mater. Today*, vol. 19, pp. 254-264, Jun. 2016.

[35] A. Mehonic *et al.*, "Silicon Oxide ($SiO_x$): A Promising Material for Resistance Switching?," *Adv. Mater.*, vol. 30, pp. 1801187, Jun. 2018.

[36] R. Bhatia, *Matrix Analysis*. New York, NY, USA: Springer-Verlag, 1997.

[37] J. R. Shewchuk, "An Introduction to the Conjugate Gradient Method without the Agonizing Pain," Carnegie Mellon University, Pittsburgh, PA, USA, Tech. Rep. CMU-CS-94-125, Aug. 1994.

[38] S. Yu *et al.*, "Scaling-up resistive synaptic arrays for neuro-inspired architecture: challenges and prospect," *IEEE International Electron Devices Meeting*, pp 17.3.1-17.3.4, Dec. 2015.

[39] International Technology Roadmap for Semiconductors (ITRS), Available at www.itrs2.net/2013-itrs.html.

[40] J. Robertson, "High dielectric constant oxides," *Eur. Phys. J. Appl. Phys.*, vol. 28, pp. 265-291, Dec. 2004.

[41] D. Heller, "A Survey of Parallel Algorithms in Numerical Linear Algebra," *SIAM Rev.*, vol. 20, pp. 740-777, 1978.

[42] L. Csanky, "Fast parallel matrix inversion algorithms," *SIAM J. Comput.*, vol. 5, pp. 618-623, 1976.

[43] V. Elango, F. Rastello, L.-N. Pouchet, J. Ramanujam, P. Sadayappan. "On characterizing the data access complexity of programs," in *Proc. of the 42nd ACM SIGPLAN-SIGACT Symposium on Principles of Programming Languages (POPL'15)*, New York, NY, USA, 2015, pp. 567-580.

[44] J. J. Hopfield, "Neural networks and physical systems with emergent collective computational abilities," *Proc. Natl. Acad. Sci. USA*, vol. 79, pp. 2554-2558, Apr. 1982.

[45] A. Cichocki, R. Unbehauen, "Neural network for solving systems of linear equations and related problems," *IEEE Trans. Circuits Syst.*, vol. 39, pp. 124-138, Feb. 1992.